\documentclass[secnumarabic,amssymb, nobibnotes, aps, pra,showpacs,preprintnumbers]{revtex4-1}
\usepackage{amssymb}
\usepackage{array}
\usepackage{graphicx}

\begin{document}

\title{Black hole temperature: minimal coupling vs conformal coupling}%

\author{Mohamadreza Fazel}
\email{m.fazel@ph.iut.ac.ir}
\affiliation{Department of Physics,
Isfahan University of Technology, Isfahan 84156-83111, Iran \\
 Departement of Physics, University of Tehran, North Kargar Avenue, Teharn 14395-547, Iran}
\author{Behrouz Mirza}
\email{b.mirza@cc.iut.ac.ir}
\author{Seyed Ali Hosseini Mansoori}
\email{sa.hosseinimansoori@ph.iut.ac.ir}
\affiliation{Department of Physics,
Isfahan University of Technology, Isfahan 84156-83111, Iran }

\begin{abstract}
In this article, we discuss the propagation of scalar fields in conformally transformed spacetimes with either minimal or conformal coupling. The conformally coupled equation of motion is transformed into a one-dimensional Schr\"{o}dinger-like equation with an invariant potential under conformal transformation. In a second stage, we argue that calculations based on conformal coupling yield the same Hawking temperature as those based on minimal coupling. Finally, it is conjectured that the quasi normal modes of black holes are invariant under conformal transformation.
\ \
\ \
\ \ \\\\\\

\end{abstract}

%%%%\pacs{04.70.Bw, 04.70.Dy, 04.62.+v}

\maketitle

\newpage
\section{Introduction}
Having incorporated the quantum behavior of a propagating field in a fixed curved background and calculated the Bogoliubov coefficients, Hawking \cite{qq11} discovered that black holes were not really black as they had been believed to be, but that they radiate with a perfect black body spectrum. Apart from Hawking's original calculation, there have been many  derivations of the Hawking effect. Euclidean signature methods have been employed to show that the Hawking temperature is related to the periodicity in imaginary time \cite{qq12}. Some authors have exploited the structure of trace anomaly to derive the Hawking radiation \cite{qq13}. Another popular version is the tunneling mechanism that obtains the Hawking flux from a path integral across the horizon \cite{qq14}. An alternative approach is to compute the reflection and absorption coefficients of quantum fields in the black hole spacetime and employing the connection between them to obtain the Bogoliubov coefficients which will ultimately yield the Hawking temperature \cite{qq7,qq8,qq9,qq10},
\begin{equation}
\frac{R}{T}=\frac{1}{{{e}^{\frac{\omega }{{{T}_{H}}}}}-1}.
\label{new19}
\end{equation}

Stephen Hawking showed that the area of event horizon never decreases in any classical physical processes \cite{SS1}. The analogy between this and the second law of thermodynamics was inspiring for Bekenstein to associate entropy to the black holes, which is proportional to the area of the event horizon \cite{SS2}. This is contrary to our intuitive expectation that the number of states is proportional to the volume of the system. Motivated by this, the holographic principle was proposed by Gerard t'Hooft \cite{SS3} and it was later developed by Leonard Susskind \cite{SS4}. It states that the information content in a macroscopic region of space can be represented by a theory that lives on the surface that encloses this region. The concrete examples of the holographic principle in the area of gravity are the black holes entropy and the AdS/CFT correspondence \cite{SS5}. The AdS/CFT correspondence is the duality between the gravity living in five dimensional AdS spacetime and the super Yang-Mills conformal field theory happening in its boundary.

The study of Hawking temperature behavior under conformal transformation is a useful method for gaining insight into the conformal invariance of such gravitational systems as black holes. No convincing evidence is available to indicate that the conformal transformation is a symmetry of  nature; however, it has been recently proposed that the conformal theory might be able to yield the evolution of the galaxies' dark matter \cite{xx1}. Moreover, a number of authors \cite{ww2} have argued that the outcomes of classical physical experiments should not be altered under such a transformation. This is contrary to the fact that this transformation does change the geometry. The Christoffel symbols and the curvature tensor are not invariant under this transformation. Besides, time like geodesics in one conformal frame is not necessarily geodesics in other conformal frames, but null rays remain geodesics although not necessarily affinely parameterized \cite{qq1,ww1}. However, since the Einstein field equation is not conformally invariant, a conformally transformed black hole solution will not equally hold for the field equations of general relativity. Nonetheless, the event horizon remains unchanged under rescaling because it is a null hypersurface and the transformed black hole may still serve as a background on which Hawking radiation  emerges and, thus, a temperature could be attributed to its event horizon. The behavior of various black holes' variables in alternative theories and for dynamical black holes under rescaling has been explored in \cite{qq3}.

In this paper, we investigate the evolution of scalar fields via both conformally  and minimally coupled equations of motion. Like the  minimal coupling equation, the conformally coupled one can  be transformed into a one-dimensional Schr\"{o}dinger-like equation with a conformally invariant potential if suitable variable changes are considered. Jacobson and Kang have argued that Hawking temperature of the asymptotically flat and static black holes remains unchanged under conformal transformations that are identity at the infinity \cite{qq2}. Here, we provide evidence in favor of the argument that the conformally coupled equation of motion gives the same Hawking temperature as does its minimally coupled counterpart and that the Hawking temperature is, therefore, invariant under the conformal transformation. We further discuss the method to identify the conformal factors that relate black holes to spacetimes with zero scalar curvature, and  examine their behavior at asymptotic infinity. Furthermore, the conformal factors are obtained in some spacetimes. The minimal coupling choice is utilized to work out the Hawking temperature for some conformally deformed black holes that are not asymptotically flat with the conformal factors that are not identity at infinity to see that the outcomes are the same as those of the original black holes. Thus, it seems that the restrictions introduced in \cite{qq2} are the sufficient conditions. Finally, it is conjectured that quasi normal modes may be invariant under conformal transformation.

The plan of this paper is as follows. The second section investigates the scalar field theory in curved spacetimes with minimal  and conformal coupling cases. In Section 3, it will be argued that both  the conformal and minimal couplings yield the same temperature for black holes. In Section 4, a differential equation will be introduced for determining the conformal factors that take us to spacetimes with zero scalar curvature. Here, it will be shown that the solutions of this differential equation go to identity at asymptotic infinity. In Section 5, it is conjectured that the Hawking temperature resulting from the minimally coupled equation is invariant under conformal transformations that are not identity at the infinity. Section 6 is devoted to the quasi normal modes. It is claimed that quasi normal modes might be invariant under conformal transformations. Conclusions and discussion are presented in Section 7. The appendices embrace brief descriptions of Whittaker  and hypergeometric functions. Additionally, we select the signature to be (-,+,+,...).

%%%%%%%%%%%%%%%%%%%%%%%%%%%%%%%%%%%%%%%%%%%%%%%%%%%%%%%%%%%%%%%%%%%%%%%%%%%%%%%%%%%%%%%%%%%%%%%%%%%%%%%%%%%%%%%%%%%%%%%%%
\section{Scalar field theory in curved spacetimes }
Either of two procedures may be exploited for generalizing the quantum field theory to curved spacetimes: minimal coupling and conformal coupling. The massless Lagrangian density for the scalar field in a curved spacetime is defined as follows
\begin{equation}
L = \sqrt { - g} ( - \frac{1}
{2}g^{\mu \nu } \nabla _\mu  \phi \,\nabla _\nu  \phi  + \xi \,R\,\phi ^2 ) \hfill \\    \hfill \\
\label{new1}
\end{equation}
where, $g$ is the determinant of the metric $g_{\mu \nu} $. A coupling to the scalar curvature $R$ has been included by the constant $\xi$. This Lagrangian density leads to the following equation of motion
\begin{equation}
\nabla _\mu  \nabla ^\mu  \phi  + \xi R \phi  = 0.
\label{new2}
\end{equation}
Minimal coupling is defined by $\xi=0$ and ignores the coupling to the curvature scalar, while in the conformal coupling it will be:
\begin{equation}
\xi  = \frac{{d - 2}}
{{4(d - 1)}}
\nonumber
\end{equation}
which is $\frac{1}{6}$ in the four dimensional spacetime. The conformally coupled equation of motion is invariant under conformal transformation
\begin{equation}
\tilde {g}_{\mu\nu}=w^{2}\left( x \right){{g}_{\mu\nu}}
\label{new3}
\end{equation}
where, the conformal factor, $w\left( x \right)$, is a smooth, non-zero, scalar function of the spacetime coordinate, if $\phi$ is also transformed as follows $\phi  \to \omega ^{\frac{d-2}{2}} \phi $. The evolution of scalar fields in curved spacetimes is described via the equation of motion in (\ref{new2}), the minimally coupled version of which is as follows and has been already  explored extensively
\begin{equation}
\nabla _\mu  \nabla ^\mu  \phi =0.
\label{new4}
\end{equation}
However, there are by far the less studies dealing with the behavior of quantum fields under the non-minimally coupled equation.  One example may be found in \cite{P1}
\begin{equation}
\nabla _\mu  \nabla ^\mu  \phi +\frac{{d - 2}}
{{4(d - 1)}}  R \phi  = 0.
\label{new5}
\end{equation}
\subsection*{Minimal coupling}

Contemplate the propagation of scalar fields via minimally coupled equation of motion (\ref{new4}) in a curved spacetime background given below
\begin{equation}
d{{s}^{2}}=-f\left( r \right)d{{t}^{2}}+\frac{d{{r}^{2}}}{f(r)}+{{r}^{2}}d\Omega _{d-2}^{2}
\label{new6}
\end{equation}
where, $d$ is the dimension of spacetime. This equation is separable by considering
 \begin{equation}
 \phi (r,t,{{\Omega }_{d-2}})=\sum\limits_{l,m}{{{e}^{i\omega t}}F(r){{Y}_{l,m}}({{\Omega }_{\text{d-2}}})},\hspace{0.5cm}F(r)=\frac{\Psi \left( r \right)}{{{r}^{^{\frac{d-2}{2}}}}}
 \label{new7}
 \end{equation}
and can be recast as a one-dimensional Schr\"{o}dinger-like equation on the interval $-\infty <x<\infty $
\begin{equation}
\frac{{{d}^{2}}\Psi (r)}{d{{x}^{2}}}+\left( {{\omega }^{2}}-V\left[ r\left( x \right) \right] \right)\Psi (r)=0
\label{new9}
\end{equation}
where, $dx=f(r)\frac{d}{dr}$ is the tortoise coordinate in which the potential depends on the geometry of the spacetime background:
\begin{eqnarray}\label{e1}
{{V}_{scalar}}\left( r \right)=f \frac{l\left( l+d-3 \right)}{{{r}^{2}}}+\\ \nonumber \frac{d-2}{2}\frac{f{f}^{'}}{r}
+\frac{\left( d-2 \right)\left( d-4 \right)}{4}\frac{f^2}{{{r}^{2}}}.
\end{eqnarray}
 A list of the potentials for some spacetimes may be found in \cite{qq4}. For the conformally transformed metric
\begin{eqnarray}\label{new12}
d\tilde s^2=-w^{2}\left( r \right)f\left( r \right)d{{t}^{2}}+\\ \nonumber \frac{w^{2}(r)}{f(r)}d{{r}^{2}}+w^{2}\left( r \right){{r}^{2}}d\text{ }\!\!\Omega\!\!\text{ }_{d-2}^{2}
\end{eqnarray}
the wave equation can also be written as
\begin{equation}
\frac{{d^2 \psi (r)}}{{dx^2 }} + (\omega ^2  - \tilde V[r(x)])\psi  (r)= 0
\label{new13}
\end{equation}
where,
\begin{equation}
\psi(r)={(rw)}^{\frac{d-2}{2}}\Psi(r)
\label{new14}
\end{equation}
 and
\begin{eqnarray}\label{new15}
\tilde V(r) = (\frac{d-2}{2})f[\frac{{w'}}{w}((\frac{d}{2} - 2)\frac{{w'}}{w}f + (d - 2)\frac{f}{r}+ \\ \nonumber
 f' ) +\frac{{w''}}{w}f + (\frac{d}{2} - 2)\frac{f}{{r^2 }} + \frac{{f'}}{r}]
+f\frac{l(l+d-3)}{r^2}
\end{eqnarray}
and $w'=\frac{d w}{dr}$. As expected, the conformal transformation changes only the potential.

   Near the horizon, both the potentials $V\left( r \right)$ and $\tilde{V}(r)$ vanish; the former is zero at the asymptotic region, and the latter vanishes at infinity in some cases. Thus the solutions of (\ref{new9}) and (\ref{new13}) will reduce to the ingoing and outgoing waves at these two regions
\begin{equation}
\Psi\left( x \right)=\psi(x)={{c}_{1}}{{e}^{i\omega x}}+{{c}_{2}}{{e}^{-i\omega x}}.
\label{new11}
\end{equation}
Various outcomes, such as absorption cross section of black holes \cite{qq5}, and Hawking temperature \cite{qq7,qq8,qq9,qq10}, can be gained by implementing different boundary conditions on the solutions to (\ref{new9}) and (\ref{new13}) at the near horizon area and at infinity.

The boundary condition for the scattering process which produces the scattering probability of waves and the absorption cross section of black hole is simple to understand \cite{qq5}. Black body radiation is produced at the black hole horizon,  part of which travels all the way to infinity, and the rest is reflected back to the black hole due to the interaction with the potential. Thus, in the near horizon we have both the ingoing and outgoing waves. However, at asymptotic infinity there is just the outgoing wave. In addition, we can consider an equivalent process which gives  similar results. An ingoing wave can be imagined at the asymptotic infinity which is scattered by the potential barrier;  there are, therefore, both ingoing and outgoing waves at infinity but only an ingoing wave  at the near horizon region.

Hawking radiation can be considered as the inverse process of scattering by the black hole. The boundary condition will, thus, involve the presence of both the ingoing and outgoing waves at the near horizon area and the absence of the outgoing wave at the infinity. The coefficients for reflection and transmission can be obtained by imposing this boundary condition. These coefficients are subsequently related to Bogoliubov coefficients, which will yield the Hawking temperature \cite{qq7,qq8,qq9,qq10}.

The reaction of a black hole perturbation will be dominated by a set of damped oscillators called quasi normal modes. They appear as solutions to  equation (\ref{new4}) when there are only ingoing waves in the horizon and outgoing waves at the infinity \cite{d1}.
\subsection*{Conformal coupling}
In order to gain a better appreciation of the scalar fields' propagation in curved spacetimes with the conformal coupling option, we recast the equation of motion (\ref{new5}) into a Schr\"{o}dinger-like equation. Then, we show that the potential is conformally invariant, and the same potential is gained in all of the conformally related spacetimes. Regarding the metric (\ref{new6})
with (\ref{new7}) and the conformally coupled equation (\ref{new5}), one can obtain
\begin{equation}
\frac{d^2 \Psi}{dx^2}+({\omega}^{2}-\hat V)\Psi =0
\label{O1}
\end{equation}
where, $\hat V$ is defined as follows
\begin{eqnarray}\label{O3}
\hat V =\frac{{d - 2}}{{d - 1}}f(\frac{{(d - 3)(d - 2)}}{{4r^2 }} - \frac{f}{{2r^2 }} + \frac{{f'}}{{2r^2 }}\\ \nonumber  - \frac{{f''}}{4})
+f\frac{l(l+d-3)}{r^2}=V-\frac{d-2}{4(d-1)}fR
\end{eqnarray}
where, $V$ is given by (\ref{e1}), and $R$ is the scalar curvature of the metric (\ref{new6}):
\begin{eqnarray}\label{O2}
R=(f'' + 2(d - 2)\frac{{f'}}{r} +\,\,\,\,\,\,\,\,\,\,\,\,\ \\ \nonumber (d - 3)(d - 2)\frac{f}{{r^2 }} - (d - 3)(d - 2)\frac{1}{r^2}).
\end{eqnarray}
Considering (\ref{new7}) and (\ref{new14}) in the conformally deformed frame (\ref{new12}),  the conformally coupled equation (\ref{new5}) can be transformed into
\begin{equation}
\frac{{d^2 \psi }}{{dx^2 }} + (\omega ^2  - \hat V)\psi  = 0
\label{new16}
\end{equation}
where, $\hat V$ is
\begin{eqnarray}\label{new18}
\hat V =\frac{{d - 2}}{{d - 1}}f(\frac{{(d - 3)(d - 2)}}{{4r^2 }} - \frac{f}{{2r^2 }} + \frac{{f'}}{{2r^2 }} - \frac{{f''}}{4})\\ \nonumber +f\frac{l(l+d-3)}{r^2}=\tilde V-\frac{(d-2)}{4(d-1)}fw^2 R_\omega \,\,\,\,\,\,\,\
\end{eqnarray}
where, $\tilde V$ is defined by (\ref{new15}), $R_w$ is the curvature scalar of metric (\ref{new12})
\begin{eqnarray}\label{new17}
R_w  = \frac{1}{{w^4 }}(2(d - 1)ww''f + (d - 4)(d - 1)w'^2 f\,\,\ \\ \nonumber +2(d - 2)(d - 1)ww'\frac{f}{r}
+2(d - 1)ww'f'+ w^2 R)
\end{eqnarray}
and $R$ is the scalar curvature (\ref{O2}).

Although the potential $\hat V$ is a function of spacetime's geometry, it is independent of the conformal factor, $w$, and is equal to the potential (\ref{O3}), as already expected. All the potentials $V$, $\tilde V$, and $\hat V$ vanish in the near horizon region, signifying that both  minimal  and conformal coupling procedures give the same physics for the near horizon vicinity  which is conformally invariant.
\section{Hawking temperature and conformal coupling}
Considering the outcomes of the former section to the effect that the scalar fields in both  conformal and minimal cases obey the same equation of motion in the near horizon vicinity, and given the well-known fact that Hawking radiation is a near horizon phenomenon, we conjecture that both minimal  and conformal coupling options, for the scalar field theory in curved spacetimes lead to the same Hawking temperature and that it will, therefore, be invariant under conformal transformation. One could calculate Hawking temperature, $T_H$, of a black hole with the metric ${g}_{\mu \nu}$ and curvature scalar $R$ using (\ref{new9}). Jacobson and Kang have argued that the temperature of asymptotically flat and static black holes is invariant under conformal transformations that are identity at infinity \cite{qq2}, which means that using (\ref{new13}) instead of (\ref{new9}) would yield the same temperature  $T_H$. Now, let us suppose that the  conformally coupled equation (\ref{O1}) is used instead of the minimally coupled counterpart  (\ref{new9}) to work out the temperature for the black hole with the metric ${g}_{\mu \nu}$, and to obtain $T'_H$. We argue that both of these equations yield the same temperatures and $T_H=T'_H$.  This will be illustrated by solving some examples.

\begin{figure}[tbp]
\centering
\fbox{\includegraphics[scale=0.3]{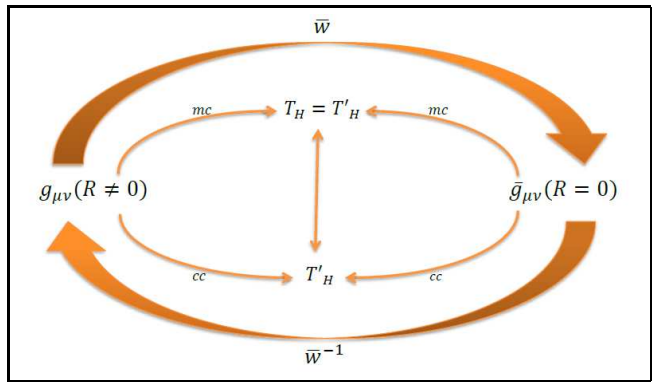} }
\caption{The spacetime with a zero scalar curvature ${\overline g}_{\mu \nu}$  which is connected to the original spacetime ${g}_{\mu \nu}$ by $\overline w(r)$ and ${\overline w(r)}^{-1}$ plays the major role in our argument.}{\label{fig1}}
\end{figure}

First, assume that we start by the conformally coupled equation (\ref{new5}) and get the temperature $T'_H$ for the black hole with the metric ${g}_{\mu \nu}$ and the curvature scalar $R$. Calculations based on this equation of motion give the same temperature for a conformally related metric due to the conformal invariance of this equation. Suppose that by choosing a suitable conformal factor, $\overline{w}$, the considered black hole spacetime is transformed into a spacetime with the metric ${\overline{g}}_{\mu \nu}=\overline{w}^{2}{g}_{\mu \nu}$ and a zero scalar curvature (Figure \ref{fig1}). It is plain that we obtain the same temperature, $T'_H$, as we would when using the conformally coupled equation (\ref{new5}) in this deformed spacetime. In such a black hole spacetime with a zero scalar curvature, there would be no difference between conformal coupling and minimal coupling and, therefore, the minimally coupled equation of motion (\ref{new4}) would lead to the Hawking temperature $T'_H$, too. Thus, both  these procedures give the same temperature $T'_H$ in the spacetime with a zero scalar curvature.
We use the inverse conformal factor, ${\overline{w}}^{-1}$, to turn back to the original black hole spacetime with the metric ${g}_{\mu \nu}={\overline{w}}^{-2} {\overline{g}}_{\mu \nu}$. Suppose that the Hawking temperature is invariant under conformal transformation ${\overline{w}}^{-1}$ under the minimal coupling option. As a result, we claim that the minimally coupled equation of motion (\ref{new4}) also yields the temperature $T'_H$ in the original spacetime. Consequently, the temperatures $T_H$ and $T'_H$ are equal. Thus we may conclude that both  (\ref{new4}) and  (\ref{new5}) yield the same temperatures for the black hole with the metric ${g}_{\mu \nu}$. Although we assume that Hawking temperature is invariant under the conformal transformation ${\overline{w}}^{-1}$ with the  minimal coupling option the non-asymptotically flat black holes' temperature and conformal transformations that are not identity at infinity violate the limitations of \cite{qq2}. In the next Section, we will show that the conformal factor ${\overline w(r)}^{-1}$ goes to identity at asymptotic infinity  for a very large group of black holes. In addition, it seems that Hawking temperature is invariant under a larger group of conformal transformations than is suggested by Jacobson and Kang \cite{qq2}. This is true even for black holes that are not asymptotically flat, as we will show through the evidence provided in Section 5 (see also \cite{P2}).
All of the above arguments are depicted in Figure \ref{fig1}, where ${g}_{\mu \nu}$, ${\overline g}_{\mu \nu}$ and $\overline{w}$ are, respectively, the metrics of the original black hole, the conformally transformed black hole with no scalar curvature, and the conformal transformation that alters the metric ${g}_{\mu \nu}$ to the metric ${\overline g}_{\mu \nu}$.  "cc" and "mc"  stand for conformal coupling and minimal coupling, respectively.

In the rest of this section, we will support our claim by employing the conformally coupled equation of motion to solve some example problems. In the case of black holes with a zero scalar curvature, such as the Schwarzschild black hole, the equivalence of minimal coupling and conformal coupling is obvious. We will work out Hawking temperatures of linear dilaton  and acoustic black holes that are not asymptotically flat. One could transform these spacetimes to one with a zero scalar curvature and a suitable conformal factor, which will be obtained in the next section. According to our arguments, the calculations based on conformally coupled equation of motion must, therefore, yield the same temperature as the one by those based on  minimal coupling. We also use the tunneling mechanism to show the same result. Although the calculations are performed in the original spacetimes, the results are also true for the conformally related black holes, due to the conformal invariance of conformally coupled equation of motion (\ref{new5}).

\subsection*{ Linear dilaton black hole}

Linear dilaton black hole is a solution to Einstein-Maxwell dilaton (EMD) theory in the four dimensional spacetime. This black hole is a special case of the more general class of non-asymptotically flat black hole solutions to EMD theory \cite{qq8,qq16}, with the following metric
 \begin{equation}
{d{s}_{1}}^2=-\frac{r-b}{r_0}dt^2+\frac{r_0}{r-b}dr^2+rr_0d\Omega ^2
\label{new44}
\end{equation}
where, $r_0$ is a constant. Taking the harmonic eigenmodes (\ref{new7}),
the radial part of the conformally coupled equation of motion (\ref{new5}) in this spacetime is
\begin{equation}
\frac{1}{r}\frac{d}{{dr}}(r(r - b)\frac{d}{{dr}} F_1(r)) + (\frac{{\tilde \omega ^2 }}{{r - b}} + \frac{{r - b}}{{12 r^2 }}) F_1(r) = 0
\label{new58}
\end{equation}
where, $\tilde\omega=r_0\omega$ and the angular momentum is zero ($l=0$). Putting $y=r-b$, (\ref{new58}) reduces to the following equation
\begin{equation}
\frac{1}{{y + b}}\frac{d}{{dy}}(y(y + b)\frac{d}{{dy}}F_1 (y)) + (\frac{{\tilde \omega ^2 }}{y} + \frac{y}{{12(y + b)^2 }})F_1(y) = 0.
\label{new59}
\end{equation}
The solution to this equation is the hypergeometric functions
\begin{eqnarray}\label{new60}
F_1 (y) = C_1 y^\beta  (y + b)^{ - \frac{{\sqrt 3 }}{6}i} F(\alpha  + \beta  + \gamma , \alpha  +\,\ \\ \nonumber \beta  - \gamma , 1 + 2\beta , - \frac{y}{b})
+ C_2 y^{ - \beta } (y + b)^{ - \frac{{\sqrt 3 }}{6}i}\,\,\,\,\,\,\ \\ \nonumber F(\alpha  - \beta  + \gamma ,\alpha  - \beta  - \gamma ,1 - 2\beta , - \frac{y}{b})\,\,\,\,\,\,\,\,\,\
\end{eqnarray}
where, $\alpha =\frac{1}{2}-\frac{\sqrt{3}}{6}i,\,\beta =i\tilde \omega $ and $\gamma =\frac{1}{6}\sqrt{6-36\tilde \omega ^2}$. In the near horizon region ($r \to b,\,\,y \to 0 $) and at infinity ($r\to \infty,\,\,y\to \infty$), the wave function $F_1(y)$ approximates to (see appendix B)
\begin{equation}
{F}_{1NH}(y)\simeq C_1{y}^{i\tilde \omega}+C_2{y}^{-i\tilde \omega}
\label{new61}
\end{equation}
\begin{eqnarray}
F_{1\infty }(y) \simeq   y^{ - \alpha  - \gamma } [ \frac{{C_1\Gamma (1 + 2\beta )\Gamma ( - 2\gamma )}}{{\Gamma (\alpha  + \beta  - \gamma )\Gamma (1 + \beta  - \alpha  - \gamma )}}+ \frac{{C_2\Gamma (1 - 2\beta )\Gamma ( - 2\gamma )}}{{\Gamma (\alpha  - \beta  - \gamma )\Gamma (1 - \beta  - \alpha  - \gamma )}}]\,\,\,\,\,\,\,\,\,\,\,\,\,\,\,\,\,\,\ \\ \nonumber
 +y^{ - \alpha  + \gamma } [ \frac{{C_1\Gamma (1 + 2\beta )\Gamma (2\gamma )}}{{\Gamma (\alpha  + \beta  + \gamma )\Gamma (1 + \beta  - \alpha  + \gamma )}}
 + \frac{{ C_2\Gamma (1 - 2\beta )\Gamma (2\gamma )}}{{\Gamma (\alpha  - \beta  + \gamma )\Gamma (1 - \beta  - \alpha  + \gamma )}}].\,\,\,\,\,\,\,\,\,\,\,\,\,\,\,\,\,\
\end{eqnarray}
The absence of an outgoing wave at infinity yields the reflection coefficient as follows
\begin{equation}
R = \left| {\frac{{\Gamma (1 - 2\beta )\Gamma (\alpha  + \beta  + \gamma )\Gamma (1 + \beta  - \alpha  + \gamma )}}{{\Gamma (1 + 2\beta )\Gamma (\alpha  - \beta  + \gamma )\Gamma (1 - \beta  - \alpha  + \gamma )}}} \right|^2 .
\label{new63}
\end{equation}
We work out (\ref{new63}) in the high frequency limit
\begin{equation}
R(\omega  \to \infty ) \simeq \frac{1}{{e^{4\pi r_0\omega }  - 1}},\, T(\omega \to \infty)=1-R\simeq 1
\label{new64}
\end{equation}
and finally, making use of (\ref{new19}), we obtain the Hawking temperature
\begin{equation}
T_H=\frac{1}{4\pi r_0}.\nonumber
\end{equation}

\subsection*{Acoustic black hole}

In 1981, Unruh pointed out the possibility of an experimental test for  black hole evaporation \cite{qq17}. He considered a nonrotational inviscid fluid with sound wave propagating in the medium. By adapting Hawking's derivation, Unruh predicted the analogue of Hawking radiation for a sound wave in a moving fluid. The linearized small perturbation of wave equations in the fluid leads to an equation of motion for the massless scalar field in a geometrical background that is similar to the black hole metric \cite{qq17}. This is the diagonalized  metric of this black hole expressed as follows:
\begin{equation}
d{{s}^{2}}=\frac{{{\rho }_{0}}}{c}\left[ -\left( {{c}^{2}}-v{{_{0}^{r}}^{2}} \right)d{{\tau }^{2}} +\frac{{{c}^{2}}d{{r}^{2}}}{{{c}^{2}}-v{{_{0}^{r}}^{2}}}+{{r}^{2}}d{{\text{ }\!\!\Omega\!\!\text{ }}^{2}} \right]
\label{new36}
\end{equation}
where, ${{\rho }_{0}},v_{0}^{r}$, and $c$ are, respectively, the density and velocity of the flow and the local velocity of the sound. If the flow smoothly exceeds the sound velocity  at the sonic horizon $r={{r}_{SH}}$, the velocity can be expanded as $v_{0}^{r}=c+a\left( r-{{r}_{SH}} \right)+O{{(r-{{r}_{H}})}^{2}}$ and, therefor, the resulting metric becomes \cite{qq9}
\begin{eqnarray}\label{new37}
d{{s}_{2}}^2={{\rho }_{0}}[ -2a\left( r-{{r}_{SH}} \right)d{{\tau }^{2}}\\ \nonumber +\frac{d{{r}^{2}}}{2a\left( r-{{r}_{SH}} \right)}+{{r}^{2}}d{{\text{ }\!\!\Omega\!\!\text{ }}^{2}}]\,\,\,\,\
\end{eqnarray}
 which is similar to the Rindler metric and the Hawking temperature is ${{T}_{H}}=\frac{a}{2\pi }$.
This black hole can also be transformed into a black hole with the zero scalar curvature (see the next section).  It is, therefore, expected that the conformal option gives the same result as the minimal one does. The radial part of the conformally coupled equation of motion (\ref{new5}) in this spacetime obeys the following equation
\begin{eqnarray}\label{A1}
\frac{1}{(z+{r}_{SH})^2}\frac{d}{dz}(kz(z+{r}_{SH})\frac{d}{dz}F_2 (z))+\,\,\ \\ \nonumber (\frac{\omega ^2}{kz}+\frac{3k(z+{r}_{SH})+k{r}_{SH}-1}{3(z+{r}_{SH})^2})F_2 (z)=0
\end{eqnarray}
where, $k=2a$ and $z=r-{r}_{SH}$. The solution to this differential equation is
\begin{eqnarray}\label{A2}
{{F}_{2}}(z)=(z+{{r}_{SH}})^{-\frac{1}{2}-\frac{{\alpha }'}{6}}({{C}_{1}}{{z}^{i\frac{\omega }{k}}}F[\frac{1}{2}-\frac{{{\alpha }'}}{6},\,\,\,\,\,\ \\ \nonumber  \frac{1}{2}-\frac{{{\alpha }'}}{6}+2i\frac{\omega }{k},1+2i\frac{\omega}{k},-\frac{z}{{{r}_{SH}}}] +{{C}_{2}}{{z}^{-i\frac{\omega }{k}}}F[\frac{1}{2}-\frac{{{\alpha }'}}{6},\,\,\,\,\ \\ \nonumber \frac{1}{2}-\frac{{{\alpha }'}}{6}-2i\frac{\omega }{k},1-2i\frac{\omega }{k},-\frac{z}{{{r}_{SH}}}])\,\,\,\,\,\,\,\,\,\,\,\,\,\,\,\,\,\,\
\end{eqnarray}
where, ${\alpha}^{'} =\frac{\sqrt{21k{r}_{SH}-12}}{\sqrt{k{r}_{SH}}}$. The behavior of (\ref{A2}) at the near horizon and asymptotic regions   are, respectively,

\begin{equation}
{F}_{2NH}(z)\simeq C_1 {z}^{i\frac{\omega}{k}}+C_2 {z}^{-i\frac{\omega}{k}}
\label{A3}
\end{equation}

\begin{eqnarray}\label{A4}
{F}_{2\infty}(z)\simeq [C_1\frac{\Gamma(1+2i\frac{\omega}{k})\Gamma(2i\frac{\omega}{k})}{\Gamma(\frac{1}{2}-\frac{{\alpha}{'}}{6}+2i\frac{\omega}{k})\Gamma(\frac{1}{2}+\frac{{\alpha}{'}}{6}+2i\frac{\omega}{k})}\,\,\ \\ \nonumber +C_2\frac{\Gamma(1-2i\frac{\omega}{k})\Gamma(2i\frac{\omega}{k})}{\Gamma(\frac{1}{2}-\frac{{\alpha}{'}}{6})\Gamma(\frac{1}{2}+\frac{{\alpha}{'}}{6})}]{z}^{1+i\frac{\omega}{k}}\,\,\,\,\,\,\,\,\,\,\,\,\,\,\,\,\,\,\,\,\,\ \\ \nonumber
+[C_1 \frac{\Gamma(1+2i\frac{\omega}{k})\Gamma(-2i\frac{\omega}{k})}{\Gamma(\frac{1}{2}-\frac{{\alpha}{'}}{6})\Gamma(\frac{1}{2}+\frac{{\alpha}{'}}{6})}\,\,\,\,\,\,\,\,\,\,\,\,\,\,\,\,\,\,\,\,\,\,\,\,\,\,\,\,\,\,\,\,\,\ \\ \nonumber +C_2\frac{\Gamma(1-2i\frac{\omega}{k})\Gamma(-2i\frac{\omega}{k})}{\Gamma(\frac{1}{2}-\frac{{\alpha}{'}}{6}+2i\frac{\omega}{k})\Gamma(\frac{1}{2}+\frac{{\alpha}{'}}{6}+2i\frac{\omega}{k})}]{z}^{1-i\frac{\omega}{k}}\,\
\end{eqnarray}
Regarding the proper boundary condition at the asymptotic region,  the reflection coefficient produced will be:
\begin{eqnarray}\label{A5}
R={\left| \frac{C_1}{C_2}\right|}^{2}=\,\,\,\,\,\,\,\,\,\,\,\,\,\,\,\,\,\,\,\,\,\,\,\,\,\,\,\,\,\,\,\,\,\,\,\,\,\,\,\,\,\,\,\,\,\ \\ \nonumber {\left| \frac{\Gamma(\frac{1}{2}-\frac{{\alpha}^{'}}{6}+2i\frac{\omega}{k})\Gamma(\frac{1}{2}+\frac{{\alpha}^{'}}{6}+2i\frac{\omega}{k})\Gamma(1-2i\frac{\omega}{k})}{\Gamma(\frac{1}{2}-\frac{{\alpha}{'}}{6})\Gamma(\frac{1}{2}+\frac{{\alpha}{'}}{6})\Gamma(1+2i\frac{\omega}{k})}\right|}^{2}.
\end{eqnarray}
In the high frequency limit, (\ref{A5}) reduces to
\begin{equation}
R(\omega \to \infty) \simeq \frac{1}{{e}^{\frac{4\pi \omega}{k}}-1},\, T(\omega \to \infty)=1-R\simeq 1
\label{A6}
\end{equation}
and it yields the Hawking temperature
\begin{equation}
T_H=\frac{k}{4\pi}=\frac{a}{2\pi}\nonumber
\end{equation}
which is like what we mentioned previously.

It should be noted that the potential in the conformal coupling option (\ref{O3}) contains two pieces, one piece is the potential of the minimal coupling case, given in (\ref{e1}) and the second part is proportional to $fR$. Due to the smoothness of the spacetime the Ricci scalar in large distances is either a constant or zero. Thus, we may consider two different behaviors for the term $fR$ at infinity. First, if the term $fR$ diverges then the potential diverges and the boundary conditions have to be handled carefully. We do not consider this case in this paper. The other possibility is when $fR$ is either zero or a constant at large distances. Thus the whole potential (\ref{O3}) is a constant now and we call it $V_0$. This case much occurs in the calculations, for example the potentials of the dilaton black hole and acoustic black holes, considered above, respectively behave as a uniform barrier and a uniform well at infinity. In this case,  equation (\ref{O1}) takes the following form at this area
\begin{equation}
\left(\frac{d^2}{dx^2}+k^2\right)\Psi(x)=0
\label{AA6}\end{equation}
where $k^2=\omega^2-V_0$.
At large distances, we have a uniform potential, however, in the large frequency limit $(\omega \rightarrow \infty)$, the propagating fields' energy is more than the potential barrier ($\omega^2-V_0>0$). Therefore, the solutions to the differential equation (\ref{AA6}) are the plane waves. In the large frequencies limit, $k$ tends to $\omega$ and the contribution of the potential is not significant, and the solutions reduce to (\ref{new11}).
Therefore, one could implement the boundary conditions without any problem in the case of asymptotically non-flat black holes when the term $fR$ is finite at asymptotic region. For the massive particles, we have an extra term containing the mass. This term is very similar to the term included the Ricci scalar, and all of the above arguments can also be applied to that.

\subsection*{Tunneling mechanism}
Another approach that yields the Hawking temperature is the tunneling mechanism. This method is based on the complex path analysis, which yield different tunneling probabilities for the ingoing and outgoing modes \cite{qq14}. This difference is as follows 
\begin{equation}
P_{emission}=e^{-\beta \omega}P_{absorption}.
\label{YY1}
\end{equation}

Considering that the potential (\ref{new18}) is vanishing in the near horizon area, the solutions to the wave equation (\ref{new16}) will be the ingoing and outgoing waves:
\begin{equation}
\psi(x)={{c}_{1}}{{e}^{i\omega x}}+{{c}_{2}}{{e}^{-i\omega x}}.
\label{k1}
\end{equation}
Transforming back to the original coordinate, it is
\begin{equation}
\psi(r)={{c}_{1}}{{e}^{i\omega \int{\frac{dr}{f(r)}}}}+{{c}_{2}}{{e}^{-i\omega \int{\frac{dr}{f(r)}}}}
\label{k2}
\end{equation}
where $f(r)$ has a pole at $r=r_H$, and the integration can be work out using the residue technique. We use the upper complex path, which detours the pole so that the pole is not inside the contour, to work out the integral when the outgoing modes locate inside the horizon ($r<r_H$) and the ingoing modes locate outside the horizon ($r>r_H$).
\begin{equation}
-\int^{r_H+\epsilon}_{r_H-\epsilon}{\frac{dr}{f(r)}}=\frac{i\pi}{f'(r_H)},
\label{k3}
\end{equation}
\begin{equation}
-\int^{r_H-\epsilon}_{r_H+\epsilon}{\frac{dr}{f(r)}}=-\frac{i\pi}{f'(r_H)}
\label{k4}
\end{equation}
where the integrals are respectively worked out for outgoing and ingoing modes and the minus sign is included to correct the sign of action \cite{qq14}. The modulus of the amplitudes give the probability of absorption and emission
\begin{equation}
P_{emission}\sim e^{\frac{-2\pi \omega}{f'(r_H)}}
\label{k5}
\end{equation}
\begin{equation}
P_{absorption}\sim e^{\frac{2\pi \omega}{f'(r_H)}}.
\label{k6}
\end{equation}
Therefore, exploiting (\ref{YY1}), we get the temperature of the black hole using the conformally coupled equation to be the same as the minimally coupled:
\begin{equation}
T=\frac{f'(r_H)}{4\pi}.
\end{equation}
We have done this calculation for all mentioned cases and confirmed that  like the method based on the absorption and transmission coefficients, the tunneling mechanism also gives the same temperature when one utilizes the conformally coupled equation of motion rather than the minimally coupled equation.

Jacobson and Kang have argued that the temperature of asymptotically flat and static black holes is invariant under conformal transformations that are identity at infinity \cite{qq2}. In this section, we argued that if a conformal factor, $\overline{w}$, exists which transforms the black hole spacetime to a spacetime with a zero scalar curvature, the minimally coupled equation of motion (\ref{new4}) and the conformally coupled equation of motion (\ref{new5}) both give the same temperature, and the temperature is, thus, invariant under any conformal transformation. Therefore, in the case of black holes with zero scalar curvature, like the case of the Schwarzschild black hole, it is, therefore, clear that the black hole's temperature is invariant under any conformal transformation. We also made use of conformal coupling to work out the Hawking temperatures of linear dilaton  and acoustic black holes, which have non-zero scalar curvatures and are conformally connected to black holes with zero scalar curvature. Besides, we show that employing the conformal coupling in the tunneling approach yields the same temperature as minimal coupling. The result was the same temperature as with the minimal coupling case confirming the argument presented in this section.

\section{Zero scalar curvature}

Generally, the conformal transformation (\ref{new3}) changes the geometry  might transform a spacetime with a non-zero scalar curvature to a spacetime with a zero scalar curvature. In this part, we will introduce a differential equation whose solutions are  appropriate conformal factors, $\overline w(r)$, that change the original spacetime into a spacetime with a zero scalar curvature. Its behavior at space-like infinity will be investigated here. This differential equation played a key role in the argument in the previous section, where we claimed that if a solution to this differential equation exists for a black hole, the minimally coupled equation (\ref{new4}) and the conformally coupled equation of motion (\ref{new5}) give the same temperature for any black holes conformally related  to it.

Assume that  the metric (\ref{new6}) is transformed under the conformal factor $\overline w(r)$ to the metric (\ref{new12}) with vanishing scalar curvature. Therefore the solutions to the following differential equation are the suitable conformal factors, which take us to the spacetimes with zero scalar curvature.
\begin{eqnarray}\label{H3}
{R}_{\overline w }=0
\end{eqnarray}
where $R_{\overline w}$ is given by (\ref{new17}). This differential equation is  non-linear  but reduces to a second order linear differential equation in the four dimensional spacetime
\begin{equation}
\overline w''(r)+(\frac{f'(r)}{f(r)}+\frac{2}{r})\overline w'(r)+\frac{R}{6f(r)}\overline w(r)=0.
\label{H4}
\end{equation}
In most cases, two independent solutions are possible to this equation employing the Frobenius method.  Now, we will examine  equation (\ref{H3}) at space-like infinity to see if the restrictions in \cite{qq2} are met. Consider the following conditions
\begin{equation}
\lim_{r\rightarrow \infty} \frac{f'(r)}{f(r)}=\frac{a}{r^n},\,\,\,\,\,\,\,\,\,\,\,\,\,\,\,\,\,\,\,\,\, \lim_{r\rightarrow \infty}\frac{f''(r)}{f(r)}=\frac{b}{r^m}
\label{H5}
\end{equation}
where, $a$ and $b$ are constants. We take $n=1$ and $m>n$, which frequently appear in this approximation. Neglecting the terms that go to zero faster than ${r}^{-1}$ in asymptotic infinity, thus $\frac{R}{f(r)}\to 0$,  we approximate the differential equation (\ref{H3}) to
\begin{equation}
\overline w''(r)+(\frac{d-4}{2})\frac{{\overline w'}^{2}}{w(r)}+(\frac{d+a-2}{r}) w'(r)=0.
\label{G1}
\end{equation}
The solution to this equation is
\begin{equation}
\overline w(r)={\left(C_1+C_2{r}^{-a-d+3}\right)}^{\frac{2}{d-2}}.
\label{G2}
\end{equation}
Another frequent possibility is $n>1$, which gives
\begin{equation}
\overline w''(r)+(\frac{d-4}{2})\frac{{\overline w'}^{2}}{w(r)}+(\frac{d-2}{r}) w'(r)=0
\label{G3}
\end{equation}
with the solution
\begin{equation}
\overline w(r)={\left(C_1+C_2{r}^{-d+3}\right)}^{\frac{2}{d-2}}
\label{G4}.
\end{equation}
The solutions (\ref{G2}) and (\ref{G4}) are  constants at asymptotic infinity, which we set to  $\overline w(r)={\overline w(r)}^{-1}=1$ for asymptotically flat black holes. The conformal transformation that changes the black hole spacetime to a spacetime with vanishing scalar curvature will, therefore, satisfy the restrictions in \cite{qq2}. These solutions in the four dimensional spacetime,
 respectively, reduce to
\begin{equation}
\overline w(r)=C_1+C_2{r}^{-(a+1)}
\label{H7}
\end{equation}
and
\begin{equation}
\overline w(r)=C_1+C_2{r}^{-1}
\label{H8}
\end{equation}
which behave the same as (\ref{G2}) and (\ref{G4}) at the asymptotic region. Moreover, it should be noted that the conditions (\ref{H5}) with the selected values for $n$ and $m$ cover a very large group of black holes, particularly asymptotically flat ones. Thus, in the case of asymptotically flat black holes, all  the restrictions in \cite{qq2} are satisfied. Based on  the results obtained in the previous  section,  we, therefore, assume that for such black holes both  the minimal  and conformal coupling options lead to  identical Hawking temperatures and  the temperature will, thus, be invariant under any conformal transformation. In the next section, we will provide some evidence indicating that the non-asymptotically flat black hole temperatures might also follow the same behavior. In the rest of this section, we solve the differential equation (\ref{H4}) for some black holes to gain the appropriate conformal transformations.

Consider the metric
\begin{equation}
d{{s}^{2}}=-k(r-b)d{{t}^{2}}+\frac{1}{k(r-b)}d{{r}^{2}}+r^2 d\Omega ^2
\label{q4}
\end{equation}
with the scalar curvature $R = 2\frac{3kr-kb-1}{r^2} $, which is similar to the acoustic black hole metric. Equation (\ref{H4}) in this spacetime takes the form
\begin{eqnarray}\label{q5}
3k(r - b)\overline w''(r) + k(9 - 6\frac{b}{r})\overline w'(r) \\ \nonumber + (\frac{{3kr - kb - 1}}{{r^2 }})\overline w(r) = 0.\,\,\,\,\,\,\,\,\
\end{eqnarray}
The Legendre functions are the solution to this equation
\begin{eqnarray}\label{q6}
\overline w(r)=\frac{C_1}{r}LegendreP[\frac{-1}{2}-ip,\frac{2b-r}{r}]\\ \nonumber +\frac{ C_2}{r}LegendreQ[-\frac{1}{2}+ip,\frac{2b-r}{r}]
\end{eqnarray}
where, $p=\frac{\sqrt{3kb+12}}{6\sqrt{kb}}$. Imagine the de Sitter spacetime, one might find various forms of this metric in \cite{S10}
\begin{equation}
ds^2=-(1-\frac{\Lambda}{3}r^2)dt^2+\frac{1}{1-\frac{\Lambda}{3}r^2}dr^2+r^2d\Omega ^2
\label{q7}
\end{equation}
with the scalar curvature $R=-4\Lambda $. The differential equation (\ref{H4}) in this spacetime is
\begin{equation}
(3-\Lambda r^2) \overline w''(r)+(\frac{6}{r}-4\Lambda r)\bar w'(r)-2\Lambda \bar w(r)=0.
\label{q8}
\end{equation}
The solution to this equation is
\begin{equation}
\overline w(r) = \frac{{C_1 }}{r} + \frac{{C_2 }}{r} {\tanh}^{-1} (\sqrt {\frac{\Lambda }{3}} r)
\label{q9}.
\end{equation}
Using (\ref{q9}) with $C_1=1$ and $C_2=0$, one can get a spacetime conformally related to (\ref{q7}) with a zero scalar curvature
\begin{equation}
d\overline s^2=-(\frac{1}{r^2}-\frac{\Lambda}{3})dt^2+\frac{1}{r^4(\frac{1}{r^2}-\frac{\Lambda}{3})}dr^2+d\Omega ^2.
\label{X1}
\end{equation}
The function "${\tanh}^{-1} (\sqrt {\frac{\Lambda }{3}} r)$" is not defined at far distances ($r\to \infty$). So, we select $C_2=0$ and then the conformal factor (\ref{q9}) will behave as (\ref{H8}) does. Nevertheless, we cannot set it to be identity at this region.

One needs to proceed with caution in calculating the temperature of the cosmological horizon in the de Sitter spacetime, because the time-like killing vector is ill-defined in some parts of this spacetime \cite{S10}, and we do not discuss this case in this paper. Nevertheless, the following naive calculations of surface gravity might indicate that the temperatures corresponding to (\ref{q7}) and (\ref{X1}) are the same. The surface gravity is given by \cite{qq1}
\begin{equation}
k=|{\frac{g'_{00}}{2\sqrt{-g_{00}g_{11}}}}|_{r=r_H}
\label{AA1}
\end{equation}
where the prime denotes differentiation with respect to the radial coordinate, $r$. Putting the elements of the metric (\ref{X1}) into the above expression for the surface gravity, we get
\begin{equation}
k=\frac{\frac{2}{r_H^3}}{2\sqrt{\frac{1}{r_H^4}}}=\frac{1}{r_H}=\sqrt{\frac{3}{\Lambda}}=2\pi T_H
\label{AA2}
\end{equation}
which is the same as the surface gravity of the original cosmological event horizon. Moreover, it can be demonstrated that the surface gravity (\ref{AA1}) is invariant under rescaling for the following general metric
\begin{equation}
d\tilde{s}^2=-\omega^2(r)f(r)dt^2+\omega^2(r)g(r)dr^2+\omega^2(r)r^2d\Omega^2
\end{equation}
where $f(r_H)=0$ and $f(r_H)g(r_H)\neq 0$. The surface gravity for the above deformed metric will be
\begin{equation}
k(\omega)=\frac{(\omega^2(r) f(r))'}{2\omega^2(r)\sqrt{f(r)g(r)}}|_{r=r_H}=\frac{f'(r)}{2\sqrt{f(r)g(r)}}|_{r=r_H}+\frac{2\omega(r)\omega'(r)f(r)}{2\omega^2(r)\sqrt{f(r)g(r)}}|_{r=r_H}.
\label{AA4}\end{equation}
The second term in the above expression vanishes, due to the root of $f(r)$ at $r=r_H$. Therefore, the surface gravity is
\begin{equation}
k(\omega)=\frac{f'(r)}{2\sqrt{f(r)g(r)}}|_{r=r_H}=k(\omega=1).
\label{AA5}\end{equation}
As we see, the surface gravity and therefore the temperature is free from the conformal factor $\omega$.

The Schwarzschild black hole has a zero scalar curvature and $\overline w$ obeys
\begin{equation}
(1-\frac{2M}{r})\overline w''(r) + \frac{2}{r}(1-\frac{M}{r})\overline w'(r) = 0
\end{equation}
with the solution
\begin{equation}
\overline w (r)= C_1 -C_2 \frac{1}{{2M}}\ln (\frac{{2M - r}}{r}).
\label{J1}
\end{equation}
The function "$\ln (\frac{{2M - r}}{r})$"  is not defined at the asymptotic region, either. We, therefore, set $C_2$ to be zero, thus (\ref{J1}) follows (\ref{H8}) at this area. We set $C_1=1$, which obeys the restriction in  \cite{qq2}. Thus, the temperature of the Schwarzschild black hole is invariant under any conformal transformation.

There are spacetimes with metrics different from (\ref{new6}). For example, consider the linear dilaton black hole (\ref{new44}).
The scalar curvature of this spacetime is $R=-\frac{r-b}{2r_0r^2}$. The conformally deformed metric can be written as
\begin{eqnarray}\label{q10}
d\overline s^2=-\overline w^2(r)\frac{r-b}{r_0}dt^2+\,\,\,\,\,\ \\ \nonumber \overline w^2(r)\frac{r_0}{r-b}dr^2+rr_0\overline w^2(r)d\Omega ^2
\end{eqnarray}
with the scalar curvature
\begin{equation}
{R}_{\overline w}=\frac{1}{2rr_0\overline w^2}(12r(r-b)\overline w''+12(2r-b)\overline w'-\overline w)=0.
\label{q11}
\end{equation}
The solution to the differential equation gives the conformal transformations that lead to the following spacetimes with no scalar curvature
\begin{eqnarray}\label{ZAZ}
\overline w(r) = C_1 F(\lambda ,\lambda ,2\lambda ,\frac{b}{r})r^{ - \lambda }  +\,\,\,\ \\ \nonumber C_2F(1 - \lambda ,1 - \lambda ,2(1 - \lambda ),\frac{b}{r})r^{\lambda  - 1}
\end{eqnarray}
where, $F(a, b, c, z)$ is a hypergeometric function (appendix B), and $\lambda=\frac{1}{2}-\frac{\sqrt{3}}{3}\simeq -0.077$. The solution (\ref{ZAZ}) is approximated  to
\begin{equation}
\overline w(r)\simeq C_1 {r}^{0.077}+C_2 {r}^{-1.077}
\end{equation}
at large distances ($r \to \infty$). We observe that it slightly disobeys (\ref{H8}) and cannot be set to be identity. However,  we saw in the previous section that the calculations based on conformally coupled equation of motion give the same temperature for this spacetime as do those based on  minimal coupling. In the next section, we will also see that the calculations based on the minimally coupled equation (\ref{new4}) in the deformed versions of this black hole leads to the same temperature. Such evidences inspire one to conjecture that the restrictions in \cite{qq2} are the only sufficient conditions.

In the third section, we argued that both the minimally coupled  and the conformally coupled equations of motion bear the same Hawking temperature if a conformal factor exists so that it changes the original black hole spacetime to one with a zero scalar curvature. In this section, we provide a differential equation for these conformal factors, and we also inspect their behavior at asymptotic infinity. As a result, we provide further confirmation for the  argument in the former section that the minimally coupled equation (\ref{new4}) and the conformally coupled equation (\ref{new5}) both lead to the same temperature for black holes. However, we conjecture that the term $\xi R$ in (\ref{new2}) is not a  factor contributing to the Hawking temperature.

\section{Conformal transformation, minimal coupling and Hawking temperature}

In this section, we   employ equation (\ref{new4}) to calculate the temperature of certain spacetimes that are conformally related to the Schwarzschild black hole, linear dilaton black hole, and acoustic black hole, for all of which we obtain the same temperature as those for the original spacetimes. Regarding the fact that linear dilaton  and acoustic black hole spacetimes are not asymptotically flat, and also given that the  conformal factors chosen are not identity at the infinity, it seems that the limitations in \cite{qq2} are sufficient conditions, and that more general conformal transformations leave the  Hawking temperature unchanged.

\subsection*{Schwarzschild black hole}

The Schwarzschild black hole with the metric
\begin{equation}
d{{s}^{2}}=-\left( 1-\frac{2M}{r} \right)d{{t}^{2}}+\frac{1}{{\left( 1-\frac{2M}{r} \right)}}d{{r}^{2}}+{{r}^{2}}d{{\text{ }\!\!\Omega\!\!\text{ }}^{2}}\,\
\label{new20}
\end{equation}
is asymptotically flat and meets one of the conditions in \cite{qq2}. The radial wave equation for the metric (\ref{new20}) is solvable in terms of the  Confluent Heun functions which are not known adequately  and one cannot, therefore,  analytically derive any outcome \cite{qq15}.

Assuming the conformal factors to be ${{\left( 1-\frac{2M}{r} \right)}^{-1}}$ and ${\left( 1-\frac{2M}{r} \right)}$,  the conformally transformed metrics will be as follows:
\begin{equation}
d{{s_3}^{2}}=-d{{t}^{2}}+{{\left( 1-\frac{2M}{r} \right)}^{-2}}d{{r}^{2}}+\frac{{{r}^{2}}}{\left( 1-\frac{2M}{r} \right)}d{{\text{ }\!\!\Omega\!\!\text{ }}^{2}}
\label{new21}
\end{equation}
\begin{equation}
d{{s_4}^{2}}=-{{\left( 1-\frac{2M}{r} \right)}^{2}}d{{t}^{2}}+d{{r}^{2}}+{{{r}^{2}}\left( 1-\frac{2M}{r} \right)}d{{\text{ }\!\!\Omega\!\!\text{ }}^{2}}
\label{new30}
\end{equation}
where the indices "3" and "4" are used, respectively, for the conformally transformed metrics under the first and  second conformal transformations. Based on (\ref{new7}), we obtain the following radial equations
\begin{equation}
{{\left( 1-\frac{2M}{r} \right)}^{2}}{{r}^{-2}}\frac{d}{dr}\left( {{r}^{2}}\frac{d}{dr} F_3 (r)\right)+{{\omega }^{2}}F_3(r)=0
\label{new23}
\end{equation}
  \begin{eqnarray}\label{new32}
  {{r}^{-2}}{{\left( 1-\frac{2M}{r} \right)}^{-2}}\frac{d}{dr}\left( {{r}^{2}}{{\left( 1-\frac{2M}{r} \right)}^{2}}\frac{dF_4(r)}{dr} \right)\\ \nonumber +{{\omega }^{2}}{{\left( 1-\frac{2M}{r} \right)}^{-2}}F_4(r)=0\,\,\,\,\,\,\,\,\,\,\,\,\,\,\,\,\,\,\,\,\,\,\,\,\
    \end{eqnarray}
where, $F_3(r)$ and $F_4(r)$, respectively, represent the radial parts of the wave function in the spacetimes (\ref{new21}) and (\ref{new30}). The scalar fields have a zero angular momentum $l=0$. Introducing $a=-2iM\omega ,b=\frac{i}{2}\sqrt{16{{M}^{2}}{{\omega }^{2}}-1}, {{b}^{'}}={}^{i\sqrt{16{{M}^{2}}{{\omega }^{2}}-9}}\!\!\diagup\!\!{}_{2}\; , z=\left( 2-\frac{r}{M} \right)$, we will have
\begin{eqnarray}\label{new24}
F_3\left( r \right)=\frac{1}{r}( {{C}_{1}}WhittakerM\left[ a,b,az \right] \\ \nonumber +{{C}_{2}}WhittakerW\left[ a,b,a\text{ }z \right] )\,\,\,\,\,\,\,\,\,\,\
\end{eqnarray}
\begin{eqnarray} \label{new34}
F_4\left( r \right)={\left( 2M-r \right)}^{-1}( {{C}_{1}}WhittakerM\left[ a,{{b}^{'}},z \right]\\ \nonumber   +{{C}_{2}}WhittakerW\left[ a,{{b}^{'}},z \right] ).\,\,\,\,\,\,\,\,\,\,\,\,\,\,\,\,\,\,\
 \end{eqnarray}
The partial wave near the horizon ($r\to 2M,z\to 0$ ) is thus (appendix A)
\begin{eqnarray}\label{new25}
{{F}_{3NH}}(r)\simeq \left( {{C}_{1}}+{{C}_{2}}\frac{\text{ }\!\!\Gamma\!\!\text{ }\left( -2b \right)}{\text{ }\!\!\Gamma\!\!\text{ }\left( \frac{1}{2}-b-a \right)} \right){{\left( az \right)}^{\frac{1}{2}+b}}\\ \nonumber +{C}_{2}\frac{\text{ }\!\!\Gamma\!\!\text{ }\left( 2b \right)}{\text{ }\!\!\Gamma\!\!\text{ }\left( \frac{1}{2}+b-a \right)}{{(az)}^{\frac{1}{2}-b}}\,\,\,\,\,\,\,\,\,\,\,\,\,\,\,\,\,\,\,\,\,\,\
\end{eqnarray}
\begin{eqnarray}
{F}_{4NH}(r)\simeq \left( {{C}_{1}}+{{C}_{2}}\frac{\text{ }\!\!\Gamma\!\!\text{ }\left( -2b' \right)}{\text{ }\!\!\Gamma\!\!\text{ }\left( \frac{1}{2}-b'-a \right)} \right){{\left( az \right)}^{\frac{1}{2}+b'}}\\ \nonumber +{{C}_{2}}\frac{\text{ }\!\!\Gamma\!\!\text{ }\left( 2b' \right)}{\text{ }\!\!\Gamma\!\!\text{ }\left( \frac{1}{2}+b'-a \right)}{{(az)}^{\frac{1}{2}-b'}}\,\,\,\,\,\,\,\,\,\,\,\,\,\,\,\,\,\,\,\,\,\,\
\end{eqnarray}
and at the infinity ($r\to \infty ,z\to \infty $), it will be:
\begin{eqnarray}\label{new26}
{{F}_{3\infty }}(r)\simeq \frac{{{C}_{1}}}{r}\frac{\text{ }\!\!\Gamma\!\!\text{ }\left( 1+2b \right)}{\text{ }\!\!\Gamma\!\!\text{ }\left( \frac{1}{2}+b-a \right)}{{e}^{iM\omega \left( \frac{r}{M}-2 \right)}}{{(az)}^{2iM\omega }}\\ \nonumber +\frac{{{C}_{2}}}{r}{{e}^{-iM\omega \left( \frac{r}{M}-2 \right)}}{{(az)}^{-2iM\omega }}\,\,\,\,\,\,\,\,\,\,\,\,\,\,\,\,\,\,\,\,\
\end{eqnarray}
\begin{eqnarray}\label{G1}
{{F}_{4\infty }}(r)\simeq \frac{{{C}_{1}}}{2M-r}\frac{\text{ }\!\!\Gamma\!\!\text{ }\left( 1+2b \right)}{\text{ }\!\!\Gamma\!\!\text{ }\left( \frac{1}{2}+b-a \right)}{{e}^{iM\omega \left( \frac{r}{M}-2 \right)}}{{(az)}^{2iM\omega }}\,\,\,\,\,\ \\ \nonumber
+\frac{{{C}_{2}}}{2M-r}{{e}^{-iM\omega \left( \frac{r}{M}-2 \right)}}{{(az)}^{-2iM\omega }}.\,\,\,\,\,\,\,\,\,\,\,\,\,\,\,\,\,\,\,\
\end{eqnarray}
Based on the relevant boundary condition, the outgoing mode at infinity must be absent and ,therefore, ${{C}_{1}}=0$; thus, the reflection coefficients are given by
\begin{equation}
R_3=\frac{{{\left| \text{ }\!\!\Gamma\!\!\text{ }\left( -2b \right)\text{ }\!\!\Gamma\!\!\text{ }\left( \frac{1}{2}+b-a \right){{\left( -2iM\omega  \right)}^{2iM\omega }} \right|}^{2}}}{{{\left| \text{ }\!\!\Gamma\!\!\text{ }\left( 2b \right)\text{ }\!\!\Gamma\!\!\text{ }\left( \frac{1}{2}-b-a \right){{\left( -2iM\omega  \right)}^{-2iM\omega }} \right|}^{2}}}
\label{new27}
\end{equation}
\begin{equation}
R_4=\frac{{{\left| \text{ }\!\!\Gamma\!\!\text{ }\left( -2b' \right)\text{ }\!\!\Gamma\!\!\text{ }\left( \frac{1}{2}+b'-a \right){{\left( -2iM\omega  \right)}^{2iM\omega }} \right|}^{2}}}{{{\left| \text{ }\!\!\Gamma\!\!\text{ }\left( 2b' \right)\text{ }\!\!\Gamma\!\!\text{ }\left( \frac{1}{2}-b'-a \right){{\left( -2iM\omega  \right)}^{-2iM\omega }} \right|}^{2}}}
\label{new27}
\end{equation}
and in the high frequency limit, both approximate to
\begin{equation}
R\left( \omega \to \infty  \right)\simeq \frac{2}{{{e}^{8\pi M\omega }}},\,\,\,\,\,\,\,\ T(\omega \to \infty)=1-R\simeq 1.
\label{new28}
\end{equation}
Putting (\ref{new28}) into (\ref{new19}) gives the Hawking temperature ($\omega \to \infty $)
\begin{equation}
{{T}_{H}}=\frac{1}{8\pi M}.
\label{new29}
\end{equation}
These two  conformal factors were identity at infinity. Below, we will consider the conformal factor ${r}^{-1}$, which is not identity at infinity and the consequent metric is given by

\begin{equation}
d{s_5}^{2}  =  - \frac{{(r - 2M)}}{{r^2 }}dt^2  + \frac{1}{{(r - 2M)}}dr^2  + rd\Omega ^2.
\label{new31}
\end{equation}
 The following equation is the radial equation in this spacetime, when $l=0$

  \begin{equation}
  \frac{d}{{dr}}((r - 2M)\frac{d}{{dr}}F_5(r)) + \frac{{r^2 w^2 }}{{(r - 2M)}}F_5(r)  = 0.
  \label{new33}
  \end{equation}
 The solution to this equation is

 \begin{eqnarray}\label{new35}
 F_5 (r) = (2M - r)^{\frac{{ - 1}}{2}} (C_1 Whittaker M[a, - a,az] \\ \nonumber + C_2 Whittaker W[a, - a,az]).\,\,\,\,\,\,\,\,\,\,\,\,\,\,\,\,\,\,\,\,\,\
  \end{eqnarray}
 The same procedure used above  leads to the Hawking temperature
 \begin{center}
  ${{T}_{H}}=\frac{1}{8\pi M}$.
 \end{center}

\subsection*{Linear dilaton black hole}

In this part, the Hawking temperature of  spacetimes that are conformally related to linear dilaton black hole spacetime (\ref{new44}) is worked   out  using the minimally coupled equation of motion (\ref{new4}). Although the conformal factors are not identity at infinity and the dilaton black hole is not asymptotically flat, the Hawking temperature remains unchanged in these conformally related spacetimes.
We select the conformal factor to be $\frac{r-b}{{{r}_{0}}}$; hence, the conformally transformed metric is
\begin{equation}
d{{s_6}^{2}}=-{{\left( \frac{r-b}{{{r}_{0}}} \right)}^{2}}d{{t}^{2}}+d{{r}^{2}}+r(r-b)d{{\text{ }\!\!\Omega\!\!\text{ }}^{2}}
\label{new45}
\end{equation}
Using (\ref{new7}) and (\ref{new45}), the radial equation takes the following form
\begin{eqnarray}\label{new46}
\frac{1}{r{{\left( r-b \right)}^{2}}}\frac{d}{dr}\left( r{{\left( r-b \right)}^{2}}\frac{dF_6(r)}{dr} \right)\,\,\,\,\,\ \\ \nonumber +\left( \frac{{{\tilde{\omega }}^{2}}}{{{\left( r-b \right)}^{2}}}-\frac{l\left( l+1 \right)}{{{r}^{2}}{{\left( r-b \right)}^{2}}} \right)F_6(r)=0
\end{eqnarray}
where, $\tilde{\omega }={{r}_{0}}\omega $. The solution to this equation is given by hypergeometric functions as follows:
\begin{eqnarray}\label{new47}
F_6(r)={{C}_{1}}{{x}^{-i\tilde{\omega }}}F\left[ \frac{1}{2}+b'-a',\frac{1}{2}-b'-a',1-2a',x \right] \nonumber \\
+{{C}_{2}}{{x}^{i\tilde{\omega }}}F\left[ \frac{1}{2}+{{b}^{'}}+{{a}^{'}},\frac{1}{2}-{{b}^{'}}+{{a}^{'}},1+2{{a}^{'}},x \right]\,\,\,\,\,\
\end{eqnarray}
where, $l=0,x=\frac{b-r}{b},{{a}^{'}}=i\tilde{\omega }$ and ${{b}^{'}}={}^{i\sqrt{16{{\tilde{\omega }}^{2}}-4}}\!\!\diagup\!\!{}_{4}\;$. Near the horizon ($x\to 0$), the solution (\ref{new47}) approximates to (appendix B)
\begin{equation}
{{F}_{6NH}(r)}\simeq {{C}_{1}}{{x}^{-i\tilde{\omega }}}+{{C}_{2}}{{x}^{i\tilde{\omega }}}.
\label{new48}
\end{equation}
And we find the asymptotic behavior ($x\to \infty$) as follows (appendix B):
\begin{eqnarray}\label{new49}
{{F}_{6\infty }(r)}\simeq [ {{C}_{1}}\frac{\text{ }\!\!\Gamma\!\!\text{ }\left( 1-2{{a}^{'}} \right)\text{ }\!\!\Gamma\!\!\text{ }\left( -2{{b}^{'}} \right)}{\text{ }\!\!\Gamma\!\!\text{ }{{\left( \frac{1}{2}-{{b}^{'}}-{{a}^{'}} \right)}^{2}}} +{{C}_{2}}\frac{\text{ }\!\!\Gamma\!\!\text{ }\left( 1+2{{a}^{'}} \right)\text{ }\!\!\Gamma\!\!\text{ }\left( -2{{b}^{'}} \right)}{\text{ }\!\!\Gamma\!\!\text{ }{{\left( \frac{1}{2}-{{b}^{'}}+{{a}^{'}} \right)}^{2}}} ]{{x}^{-\frac{1}{2}-{{b}^{'}}}}\\ \nonumber
+[ {{C}_{1}}\frac{\text{ }\!\!\Gamma\!\!\text{ }\left( 1-2{{a}^{'}} \right)\text{ }\!\!\Gamma\!\!\text{ }\left( 2{{b}^{'}} \right)}{\text{ }\!\!\Gamma\!\!\text{ }{{\left( \frac{1}{2}+{{b}^{'}}-{{a}^{'}} \right)}^{2}}}+{{C}_{2}}\frac{\text{ }\!\!\Gamma\!\!\text{ }\left( 1+2{{b}^{'}} \right)\text{ }\!\!\Gamma\!\!\text{ }\left( 2{{b}^{'}} \right)}{\text{ }\!\!\Gamma\!\!\text{ }{{\left( \frac{1}{2}+{{b}^{'}}+{{a}^{'}} \right)}^{2}}} ]{{x}^{\frac{1}{2}+{{b}^{'}}}}.\,\,\,\,\,\,\,\,\,\,\
\end{eqnarray}
The outgoing wave must be absent at the infinity; thus,
\begin{eqnarray}\label{new50}
R={{\left| \frac{{{C}_{2}}}{{{C}_{1}}} \right|}^{2}}=\frac{{{\left| \Gamma (1-2{a}')\Gamma {{(\frac{1}{2}+{b}'+{a}')}^{2}} \right|}^{2}}}{{{\left| \Gamma (1+2{a}')\Gamma {{(\frac{1}{2}+{b}'-{a}')}^{2}} \right|}^{2}}}\\ \nonumber ={{\left( {{\cos }^{2}}\left( \frac{\pi \left( 1+4i\tilde{\omega } \right)}{2} \right)-1 \right)}^{-1}}.\text{ }
\end{eqnarray}
Using (\ref{new19}) and the result (\ref{new50}) at high frequency gives:
\begin{equation}
\lim_{\omega\rightarrow \infty} \left( \frac{1}{{{e}^{\frac{\omega }{{{T}_{H}}}}}-1}\right)=\frac{1}{{{e}^{4\pi {{r}_{0}}\omega }}}.
\label{new51}
\end{equation}
We calculate the Hawking temperature as the temperature of the original spacetime
\begin{equation}
{{T}_{H}}=\frac{1}{4\pi {{r}_{0}}}.
\nonumber
\end{equation}
We also choose the conformal factor to be ${{\left( \frac{r-b}{{{r}_{0}}} \right)}^{-1}}$; so, the metric takes the following form
\begin{equation}
d{{s}_{7}}^2=-d{{t}^{2}}+{{\left( \frac{r-b}{{{r}_{0}}} \right)}^{-2}}d{{r}^{2}}+r{{\left( \frac{{{r}_{0}}}{r-b} \right)}^{2}}d{{\text{ }\!\!\Omega\!\!\text{ }}^{2}}.
\label{new52}
\end{equation}
Thus, the radial equation can be written as
\begin{eqnarray}\label{new53}
\frac{{{\left( r-b \right)}^{2}}}{r}\frac{d}{dr}\left( r\frac{dF_7(r)}{dr} \right)+\,\,\,\,\,\,\,\,\,\ \\ \nonumber \left( {{\tilde{\omega }}^{2}}-\frac{l\left( l+1 \right)}{r}{{\left( r-b \right)}^{2}} \right)F_7(r)=0.
\end{eqnarray}
Considering $l=0$, the solution to this equation is
\begin{eqnarray}\label{new54}
F_7(r)={{C}_{1}}{{x}^{\frac{1}{2}-b'}}F\left[ \frac{1}{2}-{{b}^{'}}-{{a}^{'}},\frac{1}{2}-{{b}^{'}}+{{a}^{'}},1-2{{b}^{'}},x \right]\nonumber\\
+{{C}_{2}}{{x}^{\frac{1}{2}+b'}}F\left[ \frac{1}{2}+{{b}^{'}}-{{a}^{'}},\frac{1}{2}+{{b}^{'}}+{{a}^{'}},1+2{{b}^{'}},x \right].\,\,\,\,\,\,\,\,\
\end{eqnarray}
The near horizon and asymptotic behaviors of (\ref{new54}), respectively, are (appendix B):
\begin{equation}
{{F}_{7NH}}(r)\simeq {{C}_{1}}{{x}^{\frac{1}{2}-{{b}^{'}}}}+{{C}_{2}}{{x}^{\frac{1}{2}+{{b}^{'}}}}.
\label{new55}
\end{equation}
\begin{eqnarray}\label{new56}
{{F}_{7\infty }}(r)\simeq [ {{C}_{1}}\frac{\text{ }\!\!\Gamma\!\!\text{ }\left( 1-2{{b}^{'}} \right)}{\text{ }\!\!\Gamma\!\!\text{ }{{\left( \frac{1}{2}-{{b}^{'}}+{{a}^{'}} \right)}^{2}}}+{{C}_{2}}\frac{\text{ }\!\!\Gamma\!\!\text{ }\left( 1+2{{b}^{'}} \right)}{\text{ }\!\!\Gamma\!\!\text{ }\left( \frac{1}{2}+{{b}^{'}}+{{a}^{'}} \right)} ]\text{ }\!\!\Gamma\!\!\text{ }\left( 2{{a}^{'}} \right){{x}^{{{a}^{'}}}}\,\,\,\,\,\,\,\ \\ \nonumber
+[ {{C}_{1}}\frac{\text{ }\!\!\Gamma\!\!\text{ }\left( 1-2{{b}^{'}} \right)}{\text{ }\!\!\Gamma\!\!\text{ }\left( \frac{1}{2}-{{b}^{'}}-{{a}^{'}} \right)}+{{C}_{2}}\frac{\text{ }\!\!\Gamma\!\!\text{ }\left( 1+2{{b}^{'}} \right)}{\text{ }\!\!\Gamma\!\!\text{ }\left( \frac{1}{2}+{{b}^{'}}-{{a}^{'}} \right)} ]\text{ }\!\!\Gamma\!\!\text{ }\left( -2{{a}^{'}} \right){{x}^{-{{a}^{'}}}}.\,\,\,\,\,\,\,\,\,\,\,\,\,\,\,\,\
\end{eqnarray}
Imposing a similar boundary condition at the high frequency limit gives
\begin{eqnarray}\label{new57}
R={{\left| \frac{{{C}_{2}}}{{{C}_{1}}} \right|}^{2}}=\frac{{{\left| \text{ }\!\!\Gamma\!\!\text{ }\left( \frac{1}{2}+{{b}^{'}}+{{a}^{'}} \right)\text{ }\!\!\Gamma\!\!\text{ }(1-2{{b}^{'}}) \right|}^{2}}}{{{\left| \text{ }\!\!\Gamma\!\!\text{ }\left( \frac{1}{2}-{{b}^{'}}+{{a}^{'}} \right)\text{ }\!\!\Gamma\!\!\text{ }(1+2{{b}^{'}}) \right|}^{2}}}\\ \nonumber ={{\left( {\cos}^{2} \left( \frac{\pi \left( 1+4i\tilde{\omega } \right)}{2} \right)-1 \right)}^{-1}}.
\end{eqnarray}
Obviously, the results (\ref{new50}) and (\ref{new57}) are equal; this signifies that we can obtain the same Hawking temperature as before.

\subsection*{Acoustic black hole}
We choose the conformal factor to be $1/(r{{\rho }_{0}})$; hence, the metric (\ref{new37}) takes the following form:
\begin{eqnarray}\label{new38}
d{{s_8}^{2}}=-2a\frac{(r-{r}_{SH})}{r}d{{\tau }^{2}}+\\ \nonumber \frac{1}{2ar(r-{r}_{SH})} d{{r}^{2}}+rd{{\text{ }\!\!\Omega\!\!\text{ }}^{2}}.
\end{eqnarray}
Using the separation of variables (\ref{new7}), the radial equation is written as ($l=0$):
\begin{equation}
\frac{d}{{dr}}(r(r - r_{SH} )\frac{d}{{dr}}F_8(r) ) + \frac{{\bar \omega ^2 r}}{{(r - r_{SH} )}}F_8(r)= 0
\label{new39}
\end{equation}
where, $\overline{\omega }=\frac{\omega }{2a}$. The solution to  equation (\ref{new39}) is given by the hypergeometric functions as follows:
\begin{eqnarray}\label{new40}
F_8(r)=C_1{z}^{-i\bar{\omega}}F[\alpha , \beta , \alpha +\beta , \frac{z}{{r}_{SH}}]\,\,\,\,\,\ \\ \nonumber +C_2{z}^{i\bar{\omega}}F[1-\alpha , 1-\beta , 2-(\alpha +\beta), \frac{z}{{r}_{SH}}]
\end{eqnarray}
where, $z=\left( {r}_{SH}-r \right),\alpha =\frac{1}{2}+i\sqrt{{\bar{\omega}}^{2}-\frac{1}{4}}-i\bar{\omega}$ and $\beta =\frac{1}{2}-i\sqrt{{\bar{\omega}}^{2}-\frac{1}{4}}-i\bar{\omega}$.  The behavior of (\ref{new40}) at the vicinity of horizon and at asymptotic infinity may be worked out as follows (appendix B):
\begin{equation}
{{F}_{8NH}}(r)\simeq {{C}_{1}}{{z}^{-i\overline{\omega }}}+{{C}_{2}}{{z}^{i\overline{\omega }}}\,\,\,\,\,\,\,\,\,\,\,\,\,\,\
\label{new41}
\end{equation}
\begin{eqnarray}\label{new42}
{F}_{8\infty }(r)\simeq {z}^{\frac{-1}{2}-i\bar{\omega}}[C_1\frac{\Gamma (1-2i\bar{\omega})\Gamma (-2i\bar{\omega})}{{\Gamma (\frac{1}{2}-2i\bar{\omega})}^{2}}+C_2\frac{\Gamma (1+2i\bar{\omega})\Gamma (2i\bar{\omega})}{{\Gamma (\frac{1}{2})}^{2}}]\,\,\,\,\,\,\,\,\,\,\,\,\,\,\,\,\,\,\,\,\,\,\ \\ \nonumber
+{z}^{i\bar{\omega}}[C_1\frac{\Gamma (1-2i\bar{\omega})\Gamma (2i\bar{\omega})}{{\Gamma (\frac{1}{2})}^{2}}+C_2\frac{\Gamma (1+2i\bar{\omega})\Gamma (-2i\bar{\omega})}{{\Gamma (\frac{1}{2}+2i\bar{\omega})}^{2}}]\,\,\,\,\,\,\,\,\,\,\,\,\,\,\,\,\,\,\,\,\,\,\,\,\,\,\,\,\,\,\,\,\,\,\,\
\end{eqnarray}
in which we approximate $\alpha \simeq \frac{1}{2}$ and $\beta  \simeq \frac{1}{2}-2i\bar{\omega}$ at high frequency. The outgoing wave has to be absent, which gives
 \begin{eqnarray}\label{new43}
 R={{\left| \frac{{{C}_{2}}}{{{C}_{1}}} \right|}^{2}}=\left| {\frac{{\Gamma (\frac{1}{2} + 2i\bar w)^2 }}{{\Gamma (\frac{1}{2})^2 }}} \right|^2\,\  \\ \nonumber = (1 - \cos ^2 (\frac{\pi }{2}(1 +4 i\bar \omega )))^{ - 1}.
 \end{eqnarray}
Based on (\ref{new19}) and (\ref{new43}), we may calculate the Hawking temperature at high frequency as follows
\begin{center}
${{T}_{H}}=\frac{a}{2\pi }$.
\end{center}

In summary, in this section, we have shown that Hawking temperatures of some black holes in the minimal coupling option are invariant under some conformal transformations, which are not identity at the infinity. Jacobson and Kang have claimed a similar result for the asymptotic flat black holes with the conformal factors that are identity at infinity \cite{qq2}. However, based on these samples we conjecture that  calculations using the minimally coupled equation of motion for the deformed black holes give the same temperature as the original black holes without such restrictions  (see also \cite{P2}).
\section{Quasi normal modes of liner dilaton black hole}
Quasi normal modes are the free modes of black holes. These modes are connected to the mass, spin, and charge of black holes. Their detection  could be a useful method for measuring  the parameters of black holes. In this section, we claim that these modes seem to be invariant under conformal transformation, and we provide some evidence for our hypothesis.
Considering the metric of linear dilaton black hole (\ref{new44}), the following equation may be obtained for scalar perturbations
\begin{equation}
\frac{d}{{dr}}(r(r - b)\frac{d}{{dr}}F_9(r)) + (\overline \omega  ^2 \frac{r}{{r - b}} - l(l + 1))F_9(r)= 0
\label{m1}
\end{equation}
where, $\overline \omega   = r_0 \omega $. This differential equation has the following solution \cite{qq8}:
\begin{eqnarray}\label{m2}
 F_9(r)= C_2 (\frac{{r - b}}{b})^{i\overline \omega  } F[\frac{1}{2} + i(\overline \omega   + \overline \lambda  ), \frac{1}{2}\,\,\,\,\,\,\,\,\,\,\ \\ \nonumber + i(\overline \omega   - \overline \lambda  ), 1 + 2i\overline \omega  ,\frac{{b - r}}{b}]
+ C_1 (\frac{{r - b}}{b})^{ - i\overline \omega  } F[\frac{1}{2}\,\ \\ \nonumber  - i(\overline \omega   + \overline \lambda  ), \frac{1}{2} - i(\overline \omega   - \overline \lambda  ),1 - 2i\overline \omega  ,\frac{{b - r}}{b}]\,\,\,\,\,\,\,\,\,\,\,\
 \end{eqnarray}
and $\overline \lambda  ^2  = \overline \omega  ^2  - (l + {1 \mathord{\left/
 {\vphantom {1 2}} \right.
 \kern-\nulldelimiterspace} 2})^2 $. The asymptotic behavior of this wave function at the near horizon region ($r \to b $) is \cite{qq8}:
\begin{equation}
F_{9NH} (x)\simeq C_2 e^{i\omega x}  + C_1 e^{ - i\omega x}
\label{m3}\end{equation}
where, $\frac{{r - b}}{b} = e^{{x \mathord{\left/
 {\vphantom {x {r_0 }}} \right.
 \kern-\nulldelimiterspace} {r_0 }}} $. As mentioned in Section 2,  the outgoing waves have to be absent in the near horizon vicinity if the quasi normal modes are needed to obtain. Thus, we set $C_2  = 0 $. Then, the asymptotic behavior of (\ref{m2}) at infinity becomes
\begin{eqnarray}\label{m4}
&{{F}_{9\infty }}(r)\simeq {{C}_{1}}{{(\frac{r-b}{b})}^{-\frac{1}{2}}}\Gamma (1-2i\bar{\omega })\nonumber \\
& \nonumber \times \left\{ \frac{\Gamma (2i\bar{\lambda })}{\Gamma (\frac{1}{2}-i(\bar{\omega }-\bar{\lambda }))^2}{{(\frac{r-b}{b})}^{i\bar{\lambda }}}+\frac{\Gamma (-2i\bar{\lambda })}{\Gamma (\frac{1}{2}-i(\bar{\omega }+\bar{\lambda }))^2}{{(\frac{r-b}{b})}^{-i\bar{\lambda }}} \right\}.\\
\end{eqnarray}
Removing the ingoing mode at infinity gives the quasi normal modes as follows:
\begin{equation}
\Gamma ({1 \mathord{\left/
 {\vphantom {1 2}} \right.
 \kern-\nulldelimiterspace} 2} - i(\overline \omega   + \overline \lambda  )) = \infty  \to {1 \mathord{\left/
 {\vphantom {1 2}} \right.
 \kern-\nulldelimiterspace} 2} - i(\overline \omega   + \overline \lambda  ) =  - n.
\label{m5}\end{equation}
Setting $l=0$, we work out the quasi normal modes of linear dilaton black hole:
\begin{equation}
\omega _n  = 4\pi T_H \frac{{n(n + 1)}}{{i(2n + 1)}}.
\label{m6}\end{equation}
We can calculate the quasi normal modes of deformed metrics (\ref{new45}) and (\ref{new52}). The functions (\ref{new48}) and (\ref{new49}), respectively, approximate the behavior of wave function in the near horizon region and at infinity for a spacetime defined in (\ref{new45}). To impose the boundary condition, we set $C_2=0$ in (\ref{new49}), and the absence of the ingoing mode at infinity gives
\begin{eqnarray}\label{m7}
\Gamma (\frac{1}{2} - b' - a') = \infty  \to \frac{1}{2} - i\overline \omega   - i\sqrt {\overline \omega  ^2  - \frac{1}{4}}  =  - n\,\,\,\,\,\,\,\,\,\,\  \\ \nonumber \to \omega _n  = 4\pi T_H \frac{{n(n + 1)}}{{i(2n + 1)}}.\,\,\,\,\,\,\,\,\,\,\,\,\,\,\,\,\,\,\,\,\,\,\,\,\,\,\,\,\,\,\,\,\,\,\,\,\,\,\,\,\,\,\,\,\
\end{eqnarray}
Enforcing the identical boundary conditions on (\ref{new55}) and (\ref{new56}) leads to the same quasi normal modes as (\ref{m7}). Thus, it seems that the quasi normal modes of the linear dilaton black hole are invariant under a group of conformal transformations.
\section{Discussion}
There is no insight into whether minimal  or conformal coupling is the right choice for quantum field theory in curved spacetimes. However, most authors opt for the minimal coupling option  due to its simplicity while the conformal coupling attracts less attention.

In this article, we have explored the propagation of scalar fields in curved spacetimes by both minimal  and conformal coupling options. It was shown that the minimally coupled equation of motion (\ref{new4}) in a conformally deformed spacetime could be transformed into a one dimensional Schr\"{o}dinger-like equation to derive the potential. The same was found to be true with the  conformally coupled equation (\ref{new5}). It was also demonstrated that the potential of this equation is invariant under any conformal transformation as  expected. Besides, all  these potentials vanish in the near horizon region, and it seems that both  options,  yield the same physics in the near horizon region. In \cite{qq2}, it is argued that the temperature of static and asymptotically flat black holes is invariant under rescaling with conformal factors that are identity at asymptotic infinity. \textit{Here, we conjectured that minimal scalar field theory and conformal scalar feild theory both give identical temperatures for static black holes when the term $fR$ is asymptotically finite and the black holes' spacetime is conformally related to a spacetime with no scalar curvature}. Examples were presented  to support this claim. In addition, it seems that the term $\xi R$ in (\ref{new2}) does not  contribute  to the Hawking temperature and that the  non-minimal and minimal coupling options, therefore, lead to  identical temperatures. An explanation was also put forth later as to how  the conformal factors could be found that connect the black holes to spacetimes with zero scalar curvatures and they were shown to be identity at infinity. Next, clues were provided to support the claim that the restrictions in \cite{qq2} are just the sufficient conditions. Another condition, namely, that the black hole spacetime is conformally related to a spacetime with zero scalar curvature, was explored. At last, it was  conjectured that the quasi normal modes, too, might be invariant under conformal transformations.

\section*{Appendix A: Whittaker functions}
The standard form of the Whittaker equation is
\begin{equation}
\frac{{{d}^{2}}W}{d{{z}^{2}}}+\left( -\frac{1}{4}+\frac{k}{z}+\frac{\frac{1}{4}-{{\mu }^{2}}}{{{z}^{2}}} \right)W=0
\label{a1}
\end{equation}
and the standard solutions are
\begin{equation}
WhittakerW\left[ k,\mu ,z \right],\,\,\,\,WhittakerM\left[ k,\mu ,z \right]
\label{a2}
\end{equation}
limiting forms as $z\to 0$ are \cite{qq15}
\begin{center}
$WhittakerW\left[ k,\mu ,z \right]\simeq \frac{\text{ }\!\!\Gamma\!\!\text{ }(2\mu )}{\text{ }\!\!\Gamma\!\!\text{ }\left( \frac{1}{2}+\mu -k \right)}{{z}^{\frac{1}{2}-\mu }}+\frac{\text{ }\!\!\Gamma\!\!\text{ }(-2\mu )}{\text{ }\!\!\Gamma\!\!\text{ }\left( \frac{1}{2}-\mu -k \right)}{{z}^{\frac{1}{2}+\mu }},\,\,\,\,\mu \ne 0$
\end{center}
\begin{equation}
WhittakerM\left[ k,\mu ,z \right]\simeq {{z}^{\frac{1}{2}+\mu }},\,\,\,\,\mu \ne -1,-2,\ldots
\label{a3}
\end{equation}
limiting forms as $z\to \infty $ are
\begin{eqnarray}\label{a4}
WhittakerW\left[ k,\mu ,z \right]\simeq {{z}^{k}}{{e}^{-\frac{z}{2}}},\,\,\,\,\,\,\,\,\,\,\,\,\,\,\,\,\,\,\,\,\,\ \\ \nonumber WhittakerM\left[ k,\mu ,z \right]\simeq\frac{\text{ }\!\!\Gamma\!\!\text{ }\left( \frac{1}{2}+2\text{ }\!\!\mu\!\!\text{ } \right)}{\text{ }\!\!\Gamma\!\!\text{ }\left( \frac{1}{2}+\mu -k \right)}{{z}^{-k}}{{e}^{\frac{z}{2}}}.
\end{eqnarray}
\section*{Appendix B: Hypergeometric functions}
The normal form of the hypergeometric equation is:
\begin{equation}
z\left( z-1 \right)\frac{{{d}^{2}}F}{d{{z}^{2}}}+\left( c-\left( a+b+1 \right)z \right)\frac{dF}{dz}-abF=0.
\label{b1}
\end{equation}
This equation has three singular points at $z=0,1,\infty $. The solutions in the neighborhood of these singularities are \cite{qq15}\\
singularity $z=0$
 \begin{equation}
 F\left[ a,b,c,z \right],\,\,\,\,{{z}^{1-c}}F\left[ a-c+1,b-c+1,2-c,z \right]
 \label{b2}
 \end{equation}
 singularity $z=1$
 \begin{eqnarray}\label{b3}
 F\left[ a,b,a+b-c+1,1-z \right],\,\,\,\,\,\,\,\,\,\,\,\,\,\,\,\,\,\,\,\,\,\ \\ \nonumber {{(1-z)}^{c-a-b}}F\left[ c-a,c-b,c-a-b+1,1-z \right]\,\,\,\,\,\,\
 \end{eqnarray}
 singularity $z\to \infty $
 \begin{eqnarray} \label{b4}
 {{z}^{-a}}F\left[ a,a-c+1,a-b+1,\frac{1}{z} \right],\,\ \\ \nonumber {{z}^{-b}}F\left[ b,b-c+1,b-a+1,\frac{1}{z} \right].\,\,\,\,\,\
 \end{eqnarray}
 These functions are related to each other through some linear relations as follows \cite{qq15}:
 \begin{eqnarray}\label{b5}
 F(a,b,c,z)=\frac{\Gamma (c)\Gamma (b-a)}{\Gamma (b)\Gamma (c-a)}{{(-z)}^{(-a)}}\\ \nonumber F[a,a-c+1,a-b+1,1/z]\,\,\,\,\,\,\,\,\,\,\,\,\,\,\,\,\,\,\ \\ \nonumber
 +\frac{\Gamma (c)\Gamma (a-b)}{\Gamma (a)\Gamma (c-b)}{{(-z)}^{(-b)}}\,\,\,\,\,\,\,\,\,\,\,\,\,\,\,\,\,\,\,\,\,\,\,\,\,\ \\ \nonumber F[b,b-c+1,b-a+1,1/z].\,\,\,\,\,\,\,\,\,\,\,\,\,\,\,\,\,\,\
\end{eqnarray}
\begin{eqnarray}\label{b6}
F[a,b,c,z]=\frac{\Gamma (c)\Gamma (c-a-b)}{\Gamma (c-a)\Gamma (c-b)},\,\,\,\,\,\,\,\,\ \\ \nonumber F[a,b,a+b-c+1,1-z]\,\,\,\,\,\,\,\,\,\,\,\,\,\,\,\,\,\,\,\,\,\ \\ \nonumber
+{{(1-z)}^{(c-a-b)}}\frac{\Gamma(c)\Gamma(a+b-c)}{\Gamma(a)\Gamma(b)}\,\,\,\,\ \\ \nonumber F[c-a,c-b,c-a-b+1,1-z].\,\
\end{eqnarray}
The usual normalization condition is
\begin{equation}
F\left[ a,b,c,0 \right]=1.
\label{b7}
\end{equation}
%%%%%%%%%%%%%%%%%%%%%%%%%%%%%%%%%%%%%%%%%%%%%%%%%%%%%%%%%%%%%%%%%%%%%%%%%%%%%%%%%%

\end{document}